\documentclass[jap,twocolumn,reprint,showpacs,superscriptaddress,floatfix]{revtex4-1}
\usepackage{graphicx}
\usepackage{epsfig}
\usepackage{epstopdf}
\begin{document}

\title{Effect of site occupancy disorder on Martensitic properties of Mn$_{2}$NiIn type alloys: x-ray absorption fine structure study}
\author{D. N. Lobo}
\address{Department of Physics, Goa University, Taleigao Plateau, Goa 403 206 India}
\author{K. R. Priolkar} \email[corresponding author: ]{krp@unigoa.ac.in}
\address{Department of Physics, Goa University, Taleigao Plateau, Goa 403 206 India}
\author{A. Koide}
\address{Department of Materials Molecular Science, Institute of Molecular Science, Myodaiji-cho, Okazaki, Aichi 444-8585, Japan}
\author{S. Emura}
\address{Institute of Scientific and Industrial Research, Osaka University, 8-1 Mihogaoka, Ibaraki,Osaka 567-0047, Japan}

\date{\today}

\begin{abstract}
We have carried out \textit{ab-initio} calculations of local structure of Mn and Ni in Mn$_{2}$Ni$_{1.5}$In$_{0.5}$ alloy with different site occupancies in order to understand the similarities in martensitic and magnetic properties of  Mn$_{2}$Ni$_{1+x}$In$_{1-x}$  and  Ni$_2$Mn${1+x}$In$_{1-x}$ alloys. Our results show that in Mn$_{2}$Ni$_{1+x}$In$_{1-x}$ alloys there is a strong possibility of Mn atoms occupying all the three, X, Y and Z sites of X$_2$YZ Heusler structure while Ni atoms preferentially occupy the X sites. Such a site occupancy disorder of Mn atoms is in addition to a local structural disorder due to size differences between Mn and In atoms which is also present in Ni$_2$Mn$_{1+x}$In$_{1-x}$ alloys. Further, a comparison of the calculations with experimental XAFS at the Mn and Ni K edges in Mn$_{2-y}$Ni$_{1.6+y}$In$_{0.4}$ ($-0.08 \le y \le 0.08$) indicate a strong connection between martensitic transformation and occupancy of Z sites by Mn atoms.
\end{abstract}


\maketitle

\section{Introduction}
Mn rich Ni-Mn-Z (Z = Ga, In, Sn or Sb) type shape memory alloys have been studied for their novel properties like giant reverse magneto-caloric effect \cite{krenke-natmat,preeti,ali,du,liu,khan,dub,nayak,han,sharma,sham,buchel1,li,xuan,bruno,chen,chen2}, large magnetic field induced strain \cite{kainuma,kainuma-nat,kainuma2,haldar,karaca,karaca2}, magnetic superelasticity \cite{krenke} and complex magnetic order \cite{khov,buchel2,kano,khov1,ye,kata,monroe}. The origin of all these effects lie in a strong coupling between structural and magnetic degrees of freedom. Therefore understanding the magnetic interactions between the constituent atoms as the alloys transform structurally gains importance. Despite several attempts, the understanding of the magnetism of martensitic state is still elusive. Though the magnetic moment in Ni-Mn based Heusler alloys is almost entirely due to Mn atoms\cite{webster,hurd,brown}, factors like antisite disorder\cite{felser}, changes in bond distances due to structural transformation\cite{krp-epl} as well as local structural disorder\cite{krp-prb} bring in newer magnetic interactions and add to the complexities of the problem.

Increasing the Mn content in Ni$_{2-x}$Mn$_{1+x}$Z (Z= Ga, In, Sn, Sb) alloys and at the same time preserving the Heusler structure results in alloys of type Mn$_{2}$NiZ. Band structure calculations have indicated such alloys to be ferrimagnetic  due to unequal magnetic moments of antiferromagnetically coupled Mn atoms occupying the X and Y sites of X$_2$YZ Heusler structure \cite{GDL,gdliu,Barman,aparna,Souvik}. Though martensitic transformation has been observed in Mn$_2$NiGa (T$_M$ $\sim$ 270K, T$_C$ = 588K) \cite{gdliu} the same is not observed in other Mn$_2$NiZ alloys where Z = In or Sn. However, increasing of Ni content at the expense of Z atoms to realize alloys of the type Mn$_2$Ni$_{1+x}$Z$_{1-x}$ leads to martensitic instability in them \cite{LMa,Sanchez,nelson1}.

In  L2$_1$ Heusler composition, Ni atoms are believed to prefer X sites \cite{ayuela}. According to this premise then in Mn$_2$Ni$_{1+x}$Z$_{1-x}$ alloys, if Ni atoms preferentially occupy X sites, a proportionate amount of Mn atoms would be forced to occupy Z sites leading to what is known as site occupancy disorder. Such an disorder is known to introduce competing ferromagnetic and antiferromagnetic interactions between Mn atoms occupying Y sublattice (Mn(Y)) and Mn atoms occupying Z sublattice (Mn(Z)) leading to observation of exotic properties like exchange bias effect in zero field cooled state, spin valve effect, etc. in these Mn rich martensitic alloys \cite{12singh,12nayak,tan}.

In our recent work \cite{nelson1}, the magnetic properties of Mn$_2$Ni$_{1+x}$Z$_{1-x}$ alloys in the martensitic state were found to be similar to those of Ni$_2$Mn$_{1+x}$In$_{1-x}$ and this was conjectured to be due to site occupancy disorder arising out of preferential occupation of X sites by Ni atoms. But this general picture could not explain the complete suppression of martensitic transformation in Mn$_{2-y}$Ni$_{1.6+y}$In$_{0.4}$ due to small increase in Mn concentration at the expense of Ni ($-0.1 < y < 0$) \cite{nelson1}. This is especially important because, the alloy with $y$ = 0 is martensitic with a transformation temperature of $\sim$ 230K which increases with increase in Ni concentration ($ y > 0$).  Hence it becomes necessary to understand the correlation between the perceived site occupancy disorder in Mn$_{2} $Ni$_{1+x}$In$_{1-x}$ alloys and the observed similarity in martensitic and magnetic properties of these alloys with those of Ni$_{2}$Mn$_{1+x}$In$_{1-x}$ alloys at a microscopic level.  To achieve this objective, here we report \textit{ab-initio} calculations of Ni and Mn K edge x-ray absorption fine structure (XAFS) in prototypical Mn$_{2}$Ni$_{1.5}$In$_{0.5}$ using FEFF 8.4 program and its comparison with experimental results obtained in Mn$_2$Ni$_{1.5}$In$_{0.5}$ and Mn$_{2-y}$Ni$_{1.6+y}$In$_{0.4}$ ($-0.08 \le y \le 0.08$).

\section{Methods}
The samples of above composition were prepared by arc melting the weighed constituents in argon atmosphere followed by encapsulating in a evacuated quartz tube and annealing at 750 $^\circ$C for 48 hours and subsequent quenching in ice cold water. The prepared alloys were cut in suitable sizes using a low speed diamond saw and part of the sample was powdered and re-annealed in the same procedure above. X-ray diffraction (XRD) patterns were recorded at room temperature in the angular range of $20^\circ \leq 2\theta \leq 100^\circ$ and were found to be single phase \cite{nelson1}.  Magnetization measurements were performed in the temperature interval 5 K - 400 K using a vibrating sample magnetometer in 100 Oe applied field during the zero field cooled (ZFC) and subsequent field cooled cooling (FCC) and field cooled warming (FCW) cycles.

XAFS at Ni K and Mn K edges were recorded at Photon Factory using beamline 12C at room temperature. For XAFS measurements the samples to be used as absorbers, were ground to a fine powder and uniformly distributed on a scotch tape. These sample coated strips were adjusted in number such that the absorption edge jump gave $\Delta\mu t \le 1$ where $\Delta\mu$ is the change in absorption coefficient at the absorption edge and $t$ is the thickness of the absorber. The incident and transmitted photon energies were simultaneously recorded using gas-ionization chambers as detectors. Measurements were carried out from 300 eV below the edge energy to 1000 eV above it with a 5 eV step in the pre-edge region and 2.5 eV step in the XAFS region. At each edge, at least three scans were collected to average statistical noise.

FEFF 8.4 software based on the self-consistent real-space multiple-scattering formalism \cite{Rehr} was employed for calculation of XAFS oscillations at the Mn K and Ni K edge in a prototypical alloy, Mn$_2$Ni${1.5}$In$_{0.5}$. This alloy composition is not only close to the experimentally studied compositions but also the constituent atoms have non fractional number of near neighbors. For the FEFF calculations spherical muffin tin potentials were self consistently calculated over a radius of 5\AA. A default overlapping muffin tin potentials and Hedin-Lunqvist exchange correlations were used to calculate x-ray absorption transitions to a fully relaxed final state in presence of a core hole. Calculations were carried out for Mn K and Ni K edges assuming L2$_1$ type Heusler structure. Two possible structural models and their variations which are explained in detail in next section were considered. XAFS was calculated for absorbing atoms at occupying X, Y and Z sites of the Heusler structure and combined together by multiplying each site XAFS with appropriate weighting fraction. During calculations the amplitude reduction factor, $S_0^2$ was fixed to 0.8 and the $\sigma^2$ for respective paths were calculated considering a Debye temperature of 320K \cite{chun} and the spectrum temperature of 300K

\section{Results and Discussion}
\begin{figure}
\centering
\includegraphics[width=\columnwidth]{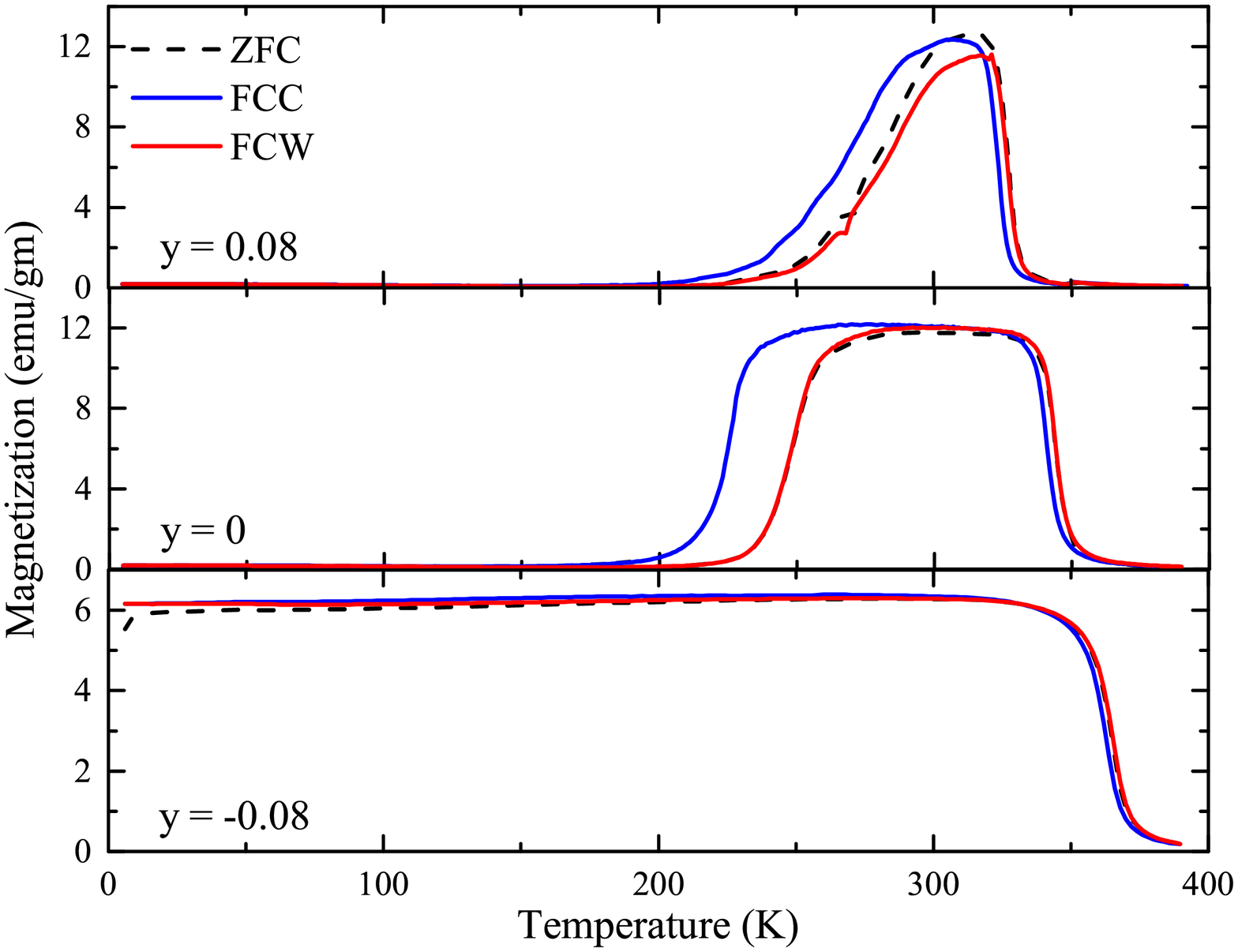}
\caption{\label{expt} Magnetization as a function of temperature for Mn$_{2-y}$Ni$_{1.6+y}$In$_{0.4}$ (y = 0.08, 0, -0.08) measured in applied field of 100 Oe during ZFC, FCC and FCW cycles.}
\end{figure}

Figure \ref{expt} presents the magnetization measurements carried out in the temperature range 5 K $\le$ T $\le$ 400K in Mn$_{2-y}$Ni$_{1.6+y}$In$_{0.4}$ (y = 0.08, 0 and -0.08). It can be clearly seen that the alloys with y = 0 and 0.08 undergo martensitic transformation at about 230 K and 270 K respectively. While Mn$_{2.08}$Ni$_{1.52}$In$_{0.4}$ does not show any martensitic instability down to 5K thus highlighting a drastic change in martensitic transformation temperature with small changes in alloy composition.  A detailed study of magnetic properties of these alloys along with Mn$_2$Ni$_{1+x}$In$_{1-x}$ (x = 0.5, 0.6 and 0.7) has been already presented in Ref. \onlinecite{nelson1}. To understand these changes in martensitic transformation temperature, experimental XAFS data recorded at the Mn K and Ni K edges at room temperature in each of these alloys have been compared with calculated Mn and Ni XAFS data using FEFF.

For the \textit{ab-initio} calculations of Mn K and Ni K edge XAFS in Mn$_2$Ni$_{1.5}$In$_{0.5}$ alloy, two structural models, designated as MODEL A and MODEL B were considered. In MODEL A, the X sites of X$_{2}$YZ are occupied equally by Ni and Mn, while all the Y sites are occupied by Mn and In and the remaining Ni atoms occupy the Z sites. In MODEL B, entire fraction of Ni atoms occupy the X sites, forcing the proportionate amount of Mn atoms to occupy the Z sites along with Y sites. Thus resulting in Mn occupying all the three X, Y and Z sites in different fractions. The site occupancies in both these models is tabulated in Table \ref{feff}.

\begin{table}
\caption{\label{feff} Assumed site occupancies of X, Y and Z sites of X$_2$YZ Heusler structure in MODEL A and MODEL B used for XAFS calculations of Mn$_2$Ni$_{1.5}$In$_{0.5}$.}
\centering
\begin{tabular}{ccccc}
\hline
Sites & MODEL A & MODEL B & MODEL B1 & MODEL B2\\
\hline
X & Ni & Ni1.5 & Ni1.5 & Ni\\
 & Mn & Mn0.5 & Mn0.5 & Mn\\
Y & Mn & Mn & Mn0.75 & Mn0.5\\
 & & & In0.25 & Ni0.5\\
Z & Ni0.5 & Mn0.5 & Mn0.75 & Mn0.5\\
 & In0.5 & In0.5 & In0.25 & In0.5\\
\hline
\end{tabular}
\end{table}

\begin{figure}
\centering
\includegraphics[width=\columnwidth]{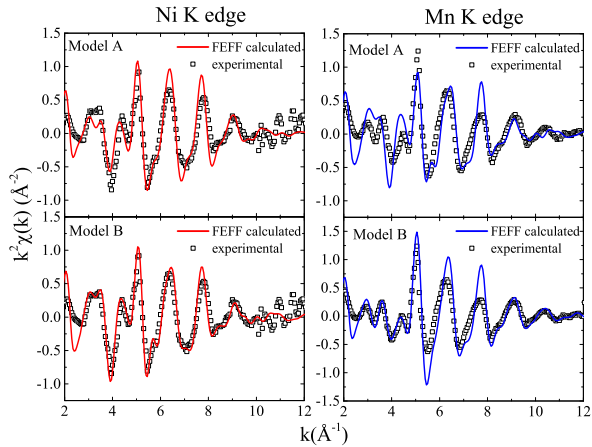}
\caption{\label{ModelA-B-Ni-Mn.eps} Calculated Ni and Mn K edge XAFS for MODEL A and MODEL B along with experimental data.}
\end{figure}

In figure \ref{ModelA-B-Ni-Mn.eps}, calculated spectra at the Ni and Mn K edges according to MODEL A and MODEL B are compared with the experimental data recorded at room temperature.  It is observed that the oscillatory parts of the experimental Mn and Ni K edge XAFS spectra are reproduced by the two theoretical models.  The calculated Ni K XAFS spectra of MODEL B  gives a much better description with the experimental data for the entire $k$ range under consideration than MODEL A. A similar conclusion could also be drawn for the calculated Mn K spectra although one can observed a mismatch between experimental and calculated Mn K XAFS spectra especially in the region between 5 to 8 \AA$^{-1}$. The in general better agreement of calculated spectra from MODEL B with the experimental XAFS spectra augers well with the literature reports that support the case of Ni atoms preferentially occupying X sites \cite{gdliu,12nayak,burch}. These reports also indicate a disorder in occupancy of Mn atoms at different sites. Such a site occupancy disorder could be the reason for observed mismatch between the experimental and calculated spectra as per MODEL B. The site disorder in occupancy of Mn atoms was introduced in the MODEL B in two different ways and are referred to as MODEL B1 and MODEL B2. Their site occupancies of the X, Y and Z sites are detailed in Table \ref{feff}.

Another reason for observed mismatch between experimental and calculated XAFS spectra could be incorrect estimation of $\sigma^2$. This is because the Debye temperature of Mn$_2$NiIn alloys have not been reported in literature and the value of 320K chosen for the present calculations is the one reported for Ni-Mn-Ga alloys. It must be mentioned here that the measured values of Debye temperature for isostructural Heusler alloys containing Mn is reported to lie between 220 K to 320 K \cite{mism} and hence the present choice of Debye temperature may not be far from the true value. Also the other extreme choice of 220K does not significantly affect the calculated spectra.

\begin{figure}
\centering
\includegraphics[width=\columnwidth]{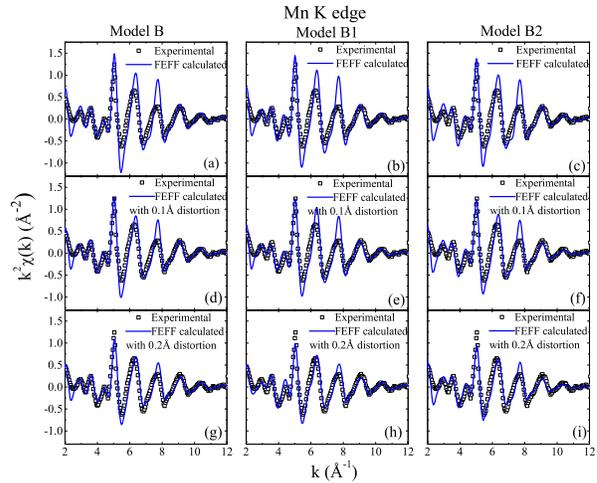}
\caption{\label{ModelB-B1-B2-Mn.eps}  Comparison of experimental XAFS data at the Mn K edge with the calculated Mn K edge XAFS for MODEL's B, B1 and B2 for undistorted lattice ((a) - (c)), for a lattice with local structural distortion wherein Mn atoms at the Z site are displaced closer to X sites by 0.1 \AA \ ((d)-(f)) and 0.2 \AA \ ((g)-(i)).}
\end{figure}

The calculated Mn K edge EXAFS of MODELS B, B1 and B2 along with the experimental data have been plotted in Fig.\ref{ModelB-B1-B2-Mn.eps}(a), Fig.\ref{ModelB-B1-B2-Mn.eps}(b) and Fig.\ref{ModelB-B1-B2-Mn.eps}(c) respectively. A comparison of the calculated Mn K edge XAFS spectra with the experimental data does not conclusively suggest any one of these models to be a better descriptor of experimental data.

In Ni$_{2}$Mn$_{1+x}$In$_{1-x} $ alloys, a local structural distortion especially in the position of Mn atoms at Z site (Mn(Z)) was shown to be responsible for the martensitic transformation for $x > 0.3$ \cite{nelson-APL}. Since in MODEL B both, Mn and In atoms occupy the Z sites, a similar local structural distortion can exist in Mn$_2$Ni$_{1+x}$In$_{1-x}$ alloys resulting in a shorter Mn(Z)-X bond as compared to In-X bond. Such a distortion was introduced by tweaking the coordinates of Mn(Z) atoms in the FEFF input file. The  coordinates of Mn(Z) atoms in MODELS B, B1 and B2 were changed in such a way that they were closer to X site atoms by 0.1 \AA \ and 0.2 \AA \ as compared to the In atoms occupying the Z sites. Fig. \ref{ModelB-B1-B2-Mn.eps}(d)-(f) and Fig. \ref{ModelB-B1-B2-Mn.eps}(g)-(i) shows the comparison for MODEL B, B1 and B2 at the Mn K edge EXAFS with experimental data for local structure distortion of 0.1\AA \ and 0.2\AA \ respectively in the k range from 2 to 12 \AA$ ^{-1} $.

\begin{figure}
\centering
\includegraphics[width=\columnwidth]{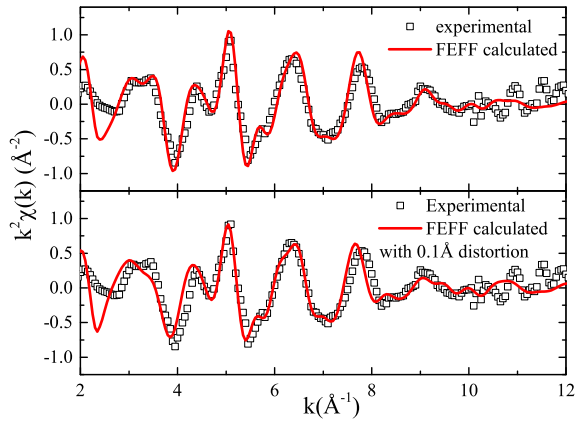}
\caption{\label{ModelB-Ni-k.eps} Ni K edge XAFS data calculated as for MODEL B for (a) undistorted lattice  and (b) lattice with local structural distortion wherein Mn atoms at In site are displaced closer to X site atoms by 0.1 \AA.}
\end{figure}

From Fig. \ref{ModelB-B1-B2-Mn.eps}, it is observed that with increasing disorder from 0.1 \AA \ to 0.2 \AA \ for all three models, the amplitude of calculated EXAFS oscillations for all models reduce and tend towards the experimental data which is an indication of presence of local structure disorder in these alloys. Since all the alloys have long range structural order as evidenced from Bragg reflections in x-ray diffraction data, a displacement of a particular atom by 0.2\AA \ from its crystallographic site position may be a bit unrealistic. Hence models with 0.1\AA \ displacement of Mn(Z) atoms were taken to provide the most realistic description of site occupancies in such Mn$_2$Ni$_{1+x}$In$_{1-x}$ alloys. Of the three MODEL B was preferred over MODELS B1 and B2 due to its relative simplicity. Irrespective of the choice of models, the present analysis clearly suggests antisite disorder along with local structural disorder to be primarily responsible for physical properties of  Mn$_{2}$NiIn type alloys. Antisite disorder has also been reported to be responsible for exotic properties like spin valve effect and zero field exchange bias in related Mn$_{2} $NiGa and Mn$_{2} $PtGa \cite{12singh,12nayak}.

Calculated Ni K edge XAFS plot for MODEL B with a local structure distortion of  0.1 \AA \  also presents a better agreement with the experimental data as compared to the undistorted MODEL B (See Fig. \ref{ModelB-Ni-k.eps}). This confirms the presence of antisite disorder along with a local structural disorder in these Mn$_{2}$NiIn  type alloys. Presence of Mn at the Z sites could be the reason for similarity of magnetic properties in martensitic state of Mn$_{2}$Ni$_{1+x}$In$_{1-x}$ and the magnetic properties of Ni$_{2}$Mn$_{1+x}$In$_{1-x}$ alloys in their martensitic state. In other words, an increase in Ni content at the expense of In in Mn-Ni-In alloys  causes  Mn to occupy the Z sites due to preference of Ni atoms for X sites. Such a site occupancy coupled with local structural disorder favors formation of  Ni-Mn(Z) hybridization which is responsible for martensitic transformation and antiferromagnetism in the martensitic state.

Although above calculations give a fair understanding of magnetic properties of the martensitic state in Mn$_2$NiIn type alloys, it does not give explain the complete suppression of martensitic transformation with small changes in Mn:Ni ratio. In Mn$_{2}$Ni$_{1+x}$In$_{1-x}$, though alloy with $x = 0.6$ undergoes martensitic transformation at 232K, the alloy Mn$_{2.08}$Ni$_{1.52}$In$_{0.4}$ alloy does not exhibit any martensitic transformation down to 4K. At the same time Mn$_{1.92}$Ni$_{1.68}$In$_{0.4}$ exhibits martensitic transformation at a higher temperature of 250K. To understand the possible cause of such drastic variation of martensitic transformation temperatures in these alloys, a linear component fitting (LCF) analysis was performed on the experimental data recorded at the Mn and Ni K edge XAFS in the above three compositions using the FEFF calculated XAFS of Ni and Mn occupying different site position as per MODEL A and MODEL B with distortion of 0.1\AA. As per MODEL A, Ni would be found at X and Z sites while Mn would be present at X and Y sites. In case of MODEL B, Ni would be present only at X sites and Mn would occupy all the three sites. Calculated XAFS of Ni and Mn for each site were used as standards and Athena was employed to give a best possible combination that describes the experimental data in Mn$_{2-y}$Ni$_{1.6+y}$In$_{0.4}$. Based on this LCF analysis the obtained site occupancies of Ni and Mn are presented in Table \ref{LCF-Ni} and Table \ref{LCF-Mn} respectively.

\begin{table}
\caption{\label{LCF-Ni}LCF analysis for Ni K edge XAFS The bracketed letters indicates the crystallographic site positions.}

\begin{tabular}{llll}
\hline Sample & \multicolumn{3}{c}{Ni EXAFS}\\

& Model     & Ni-Mn(Z)  & species  \\
&  (position)     & bond distance  & concentration \\
&                 & disorder  &  \\
\hline
Mn$_{2.08}$Mn$_{1.52}$In$_{0.4}$ &   A(X) & -  &  79 $\pm$ 20 \\
                           &   A(Z) & -  & 14 $\pm$ 07    \\
                           &   B(X) & 0.1 & 12 $\pm$ 14  \\
Mn$_{2}$Ni$_{1.6}$In$_{0.4}$ &   A(X) &  - &  83 $\pm$ 13 \\
                                 &   B(X) & 0.1 & 17 $\pm$ 11  \\
Mn$_{1.92}$Ni$_{1.68}$In$_{0.4}$ &   A(X) &  - &  86 $\pm$ 14\\
                                 &   B(X) & 0.1 & 14 $\pm$ 13  \\
\hline
\end{tabular}
\end{table}

\begin{table}
\caption{\label{LCF-Mn}LCF analysis for Mn K edge XAFS The bracketed letters indicates the crystallographic site positions.}
\begin{tabular}{llll}
\hline Sample & \multicolumn{3}{c}{Mn EXAFS}\\

& Model & Ni-Mn(Z)  & species  \\
&  (position)       & bond distance  & concentration \\
&                   & disorder           &    \\
\hline
Mn$_{2.08}$Ni$_{1.52}$In$_{0.4}$ &   A(X) &  - &  62 $\pm$ 05 \\
                                 &   B(Y) & 0.1 & 38 $\pm$ 05  \\
Mn$_2$Mn$_{1.6}$In$_{0.4}$ &   A(X) & -  &  49 $\pm$ 6 \\
                           &   B(Y) & 0.1 & 31 $\pm$ 10    \\
                           &   B(Z) & 0.1 & 20 $\pm$ 07  \\
Mn$_{1.92}$Ni$_{1.68}$In$_{0.4}$ &   A(X) &  - &  50 $\pm$ 25\\
                                 &   B(Y) & 0.1 & 23 $\pm$ 5.4  \\
                                 &   B(Y) & 0.1 & 27 $\pm$ 7.7 \\
\hline
\end{tabular}
\end{table}

It is interesting to note that in case of Mn$_{2.08}$Ni$_{1.92}$In$_{0.4}$ alloy which does not undergo martensitic  transformation, Mn is found to be only at X and Y sites and while Ni is present at the Z site. While in case of the other two alloys, Ni primarily occupies X sites while Mn is found to occupy all the three sites. Presence of  Mn at the Z site along with a local structural distortion in its position gives rise to a shorter Ni-Mn bond as compared to Ni-In and Ni(3d) - Mn(3d) hybridization which plays an important role in martensitic transformation in Ni$_2$Mn$_{1+x}$In$_{1-x}$ alloys. A clear differentiation between site occupancies of Mn$_2$NiIn type alloys undergoing martensitic transformation and non-martensitic alloys highlights the importance of antisite disorder along with local structural distortion in inducing martensitic transformation in these alloys.

\section{Conclusion}
We have carried out ab-initio calculations at the Ni and Mn K edge to understand the driving force for martensitic transformation in Mn$_{2}$Ni$_{1+x}$In$_{1-x}$ alloys. Presence of Mn at Z sites appears to be the main requirement for the alloy composition to undergo martensitic transformation. The ab-initio XAFS calculations indicate preferential occupation of X sites by Ni atoms while Mn occupy all X, Y and Z sites of the X$_{2} $YZ Heusler structure. Such a site occupancy disorder of Mn atoms is in addition to a local structural disorder due to size differences between Mn and In atoms which is also present in Ni$_2$Mn$_{1+x}$In$_{1-x}$ alloys. This augers well with the observed similarities in magnetic properties of martensitic state of  Mn$_{2}$Ni$_{1+x}$In$_{1-x}$ and  Ni$_2$Mn$_{1+x}$In$_{1-x}$ alloys. Further the drastic suppression of martensitic transformation with small changes in composition in Mn$_{2-y}$Ni$_{1.6+y}$In$_{0.4}$ can also be understood based on occupancy of Mn at Z sites. Mn$_{2.08}$Ni$_{1.52}$In$_{0.4}$ which has no Mn atoms at the Z site, does not undergo martensitic transformation while Mn$_2$Ni$_{1.6}$In$_{0.4}$ which has about 20\% Z site occupancy of Mn, undergoes martensitic transformation at about 230K.

\section*{Acknowledgements}
Authors would like to acknowledge the financial assistance from Science and Engineering Research Board (SERB), DST, Govt. of India under the project SB/S2/CMP-096/2013. The work at Photon Factory was performed under the Proposal No. 2011G0077

\bibliographystyle{apsrev4-1}
\bibliography{Mn2NiIn-abinitio}

\end{document}